\documentstyle[11pt,epsf]{article}
\def\GeV{\,{\rm GeV}}
\def\TeV{\,{\rm TeV}}

\def\sec{\,{\rm sec}}

\def\rcm{\,{\rm cm}}
\def\km{\,{\rm km}}

\def\Mpc{\,{\rm Mpc}}

\def\eV{{\,\rm eV}}

\def\cmm2{{\,\rm cm^{-2}}}
\def\cm2{{\,{\rm cm}^2}}
\def\cmm3{{\,{\rm cm}^{-3}}}
\def\gcmm3{{\,{\rm g\,cm^{-3}}}}
\def\kms{\,{\rm km\,s^{-1}}}

\def\mpl{{m_{\rm Pl}}}

\def\la{\mathrel{\mathpalette\fun <}}
\def\ga{\mathrel{\mathpalette\fun >}}
\def\fun#1#2{\lower3.6pt\vbox{\baselineskip0pt\lineskip.9pt
  \ialign{$\mathsurround=0pt#1\hfil##\hfil$\crcr#2\crcr\sim\crcr}}}
\begin{document}
\newpage
\pagestyle{plain}
\setcounter{page}{1}

\begin{center}
\bigskip

%\rightline{FERMILAB--Pub--95/090-A}
%\rightline{astro-ph/9703194}

{\LARGE \bf Inflation:  From Theory\\
\medskip
To Observation and Back}

\vspace{22pt}
MICHAEL S. TURNER\\

\vspace{11pt}

{\it Departments of Physics and of Astronomy \& Astrophysics\\
Enrico Fermi Institute, The University of Chicago, Chicago, IL~~60637-1433}\\

\smallskip
{\it and}
\smallskip

{\it NASA/Fermilab Astrophysics Center\\
Fermi National Accelerator Laboratory, Batavia, IL~~60510-0500}\\

\end{center}

\section*{OVERVIEW}

Alan Guth introduced cosmologists to inflation at the 1980
Texas Symposium.  Since, inflation has had almost as much impact
on cosmology as the big-bang model itself.  However, unlike
the big-bang model, it has little observational support.
Hopefully, that situation is about to change as a variety
and abundance of data begin to test inflation in a significant
way.  The observations that are putting inflation to test involve
the formation of structure in the Universe, especially measurements
of the anisotropy of the cosmic background radiation.  The
cold dark matter models of structure formation motivated by
inflation are holding up well as the observational tests become
sharper.  In the next decade inflation will be tested even more
significantly, with more precise measurements of CBR anisotropy,
the mean density of the Universe, the Hubble constant, and the distribution of
matter, as well as sensitive searches for the nonbaryonic dark
matter predicted to exist by inflation.
As an optimist I believe that we may be well on our way to a standard
cosmology that includes inflation and extends back to around
$10^{-32}\sec$, providing an important window on the earliest
moments and fundamental physics.

\section{BEYOND THE BIG BANG MODEL}

The hot big-bang cosmology is a remarkable achievement.  It
provides a reliable account of the Universe from about $10^{-2}\sec$
to the present.  Further, it together with modern ideas in
particle physics---the Standard Model, supersymmetry, grand
unification, and superstring theory---provides a sound framework
for sensible speculation all the way back to the Planck epoch
and perhaps even earlier.\footnote{Before the advent of the Standard
Model (point-like quarks and leptons
with ``weak interactions'' at short distances)
cosmology hit a wall at about $10^{-5}\sec$.  At this time
the Universe was a strongly interacting gas of overlapping hadrons
with the number of ``fundamental particles'' increasing
exponentially with mass.}

These speculations have allowed cosmologists to address
a deeper set of questions:  What is the
nature of the ubiquitous dark matter that is the dominant component
of the mass density?  Why does the Universe contain only matter?
What is the origin of the tiny inhomogeneities
that seeded the formation of structure, and how did that structure
evolve?  Why is the portion of the Universe that we can see so
flat and smooth?  What is the value of the vexing cosmological
constant?  How did the expansion begin---or was there a beginning?

In the past fifteen years much progress has been made, and
many believe that the answers to all these questions involve
events that took place during the earliest moments and involved physics
beyond the Standard Model \cite{eu}.  For example, the matter-antimatter asymmetry,
quantified as a net baryon number of about $10^{-10}$ per photon, is
believed to have developed through interactions that do not conserve
baryon number and $C$, $CP$ (matter-antimatter symmetry) and
occurred out of thermal equilibrium.  Until recently it was believed
that ``baryogenesis'' involved unification-scale physics and occurred
around $10^{-34}\sec$; recent work suggests that baryogenesis
might have occurred at the weak scale ($T\sim 300\GeV$ and
$t\sim 10^{-11}\sec$) and involved
baryon-number and $C$, $CP$ violation within the
Standard Model \cite{ewbaryo}.

The most optimistic early-Universe cosmologists (of which I am one)
believe that we are on the verge of solving all of the above problems
and extending our knowledge of the Universe back to around $10^{-32}
\sec$ after ``the bang.''  The key to this is inflation.  Among other
things, inflation has led to the cold dark matter models of structure formation,
which are characterized by scale-invariant density perturbations and
dark matter whose composition is primarily slowly moving elementary
particles (e.g., axions or neutralinos).   The cold dark matter
theory is crucial to testing inflation,
and if it proves correct, would complete the standard cosmology
by connecting the theorist's early Universe which is smooth
and formless to the astronomer's Universe which is
inhomogeneous and abounds with structure.

\subsection{Evidence}

Four pillars provide the observational support on which the
hot big-bang model rests:  (1) The uniform distribution of matter
on large scales and the isotropic expansion that maintains this
uniformity; (2) The existence of a nearly uniform and accurately
thermal cosmic background radiation (CBR); (3) The abundances
(relative to hydrogen) of the light elements D, $^3$He, $^4$He,
and $^7$Li; and (4) The existence of small fluctuations in the
temperature of the CBR across the sky at the level about $10^{-5}$.
The Hubble expansion supports the general notion of an expanding
Universe; the CBR provides almost indisputable evidence of a
hot, dense beginning.  The agreement between the light-element
abundances predicted by primordial nucleosynthesis and those
observed tests the model back to about $10^{-2}\sec$ and
leads to the most accurate determination of the baryon density,
$\Omega_B \simeq 0.009h^{-2} - 0.022h^{-2}$ \cite{copi}.
The small fluctuations in the temperature of CBR across the
sky indicate the existence of primeval density perturbations
of a similar size which amplified by gravity over the age
of the Universe has led to the abundance of structure seen today (galaxies,
clusters of galaxies, superclusters, voids, and great walls).

At present, observational support for inflation is fragmentary at
best.  However, a wealth of diverse observations are beginning to
seriously test inflation, and even someone who is not an early-Universe
optimist would have to concede that inflation is likely to be tested
in a significant way within the next five years or so.  The crucial tests
include measurements of CBR anisotropy, the present degree of inhomogeneity
as probed by redshift surveys and peculiar-velocity measurements,
x-ray studies of clusters of galaxies, increasingly accurate measurements
of the Hubble constant, the study of the Universe at high redshift
by large ground-based telescopes and the Hubble Space Telescope,
the mapping of dark matter through gravitational lensing,
the search for baryonic dark matter through
microlensing and particle dark matter through both direct and indirect
techniques, and on and on.

While the observational support for inflation is not
overwhelming---yet!---there
are a number of observations which are very encouraging.  The
evidence that the mass density of the Universe is significantly
larger than that which baryons can account for continues to grow:
measurements based upon peculiar velocities of the Milky Way
and other galaxies indicate that $\Omega_{\rm matter} \ga 0.3$
\cite{dekel,davis} and x-ray and weak-gravitational lensing measurements
of the masses of rich clusters continue to indicate mass to light
ratios consistent with $\Omega_{\rm matter} \ga 0.2$ or greater.  (For
a Hubble constant of greater than $60\kms\Mpc^{-1}$, baryons
can contribute at most 5\% of the critical density.)   Further,
the mass fraction of rich clusters that can be readily identified
as baryonic (mostly hot gas) is only about $0.04h^{-3/2} - 0.1 h^{-3/2}$
\cite{clusters}.  Likewise, the fraction of the dark halo of our galaxy that
can be accounted for by baryons (faint stars and MACHOs) is only
about $0.05 - 0.3$ \cite{machofrac}.  CBR anisotropy
has now been measured on angular scales from about $0.5^\circ$ to
$90^\circ$ \cite{wss,wil}, probing the spectrum of metric perturbations
on scales from about $30h^{-1}\Mpc$ to almost $10^4h^{-1}\Mpc$.
The measurements are consistent with the scale-invariant prediction
of inflation; see Fig.~1.  Likewise, measurements of the spectrum of
inhomogeneity on smaller scales from redshift surveys, say less than about
$100h^{-1}\Mpc$, are generally consistent with the predictions of
cold dark matter model (see Fig.~2).  The success of cold dark matter
is all the more striking given that the only other models (Peebles'
baryons only PBI model and topological defects + nonbaryonic dark
matter) are on the verge of being ruled out.

\begin{figure}[htb]
\centering
\epsfxsize=12cm
\leavevmode
{\epsfbox{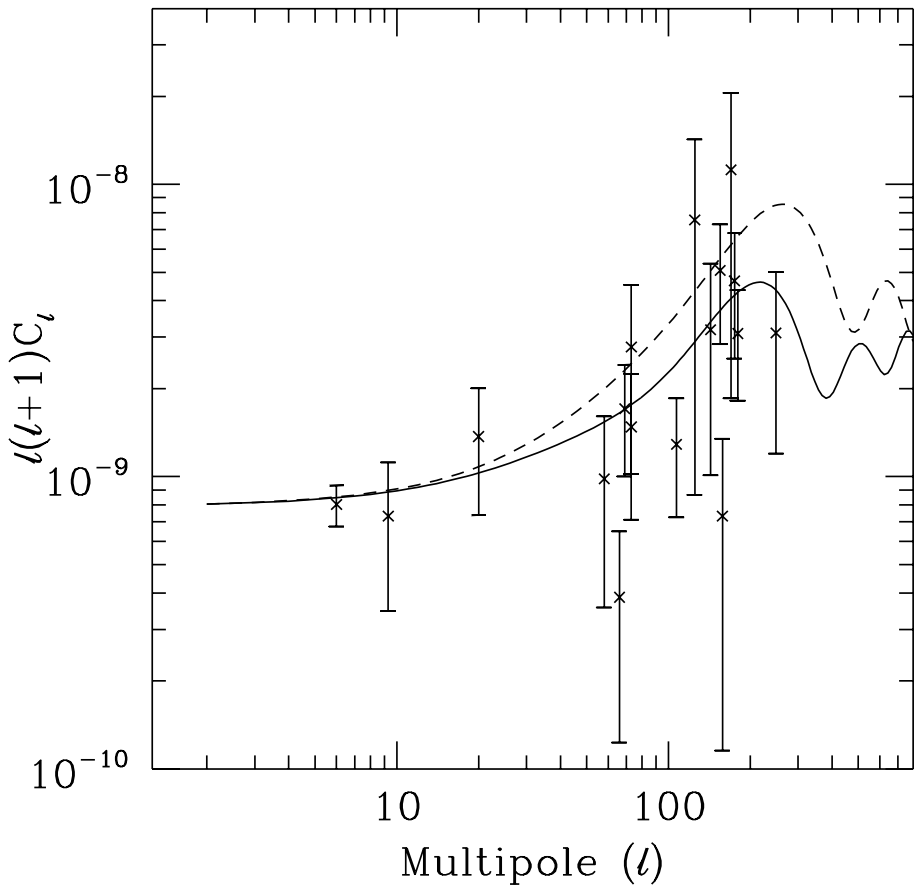}}
\caption{
{Summary of current measurements of CBR
anisotropy in terms of a spherical-harmonic decomposition,
$C_l \equiv \langle |a_{lm}|^2\rangle$.
The rms temperature fluctuation measured between two
points separated by an angle $\theta$
is roughly given by:  $(\delta T/T)_{\theta} \simeq
\protect\sqrt{l(l+1)C_l}$ with $l \simeq 200^\circ /\theta$.
The curves are the cold dark matter predictions,
normalized to the COBE detection, for Hubble constants
of $50\kms\Mpc$ (solid) and $35\kms\Mpc^{-1}$ (broken).
(Figure courtesy of M.~White.)}
  }
\label{fig:1}
\end{figure}

\begin{figure}[htb]
\centering
\epsfxsize=13cm
\leavevmode
{\epsfbox{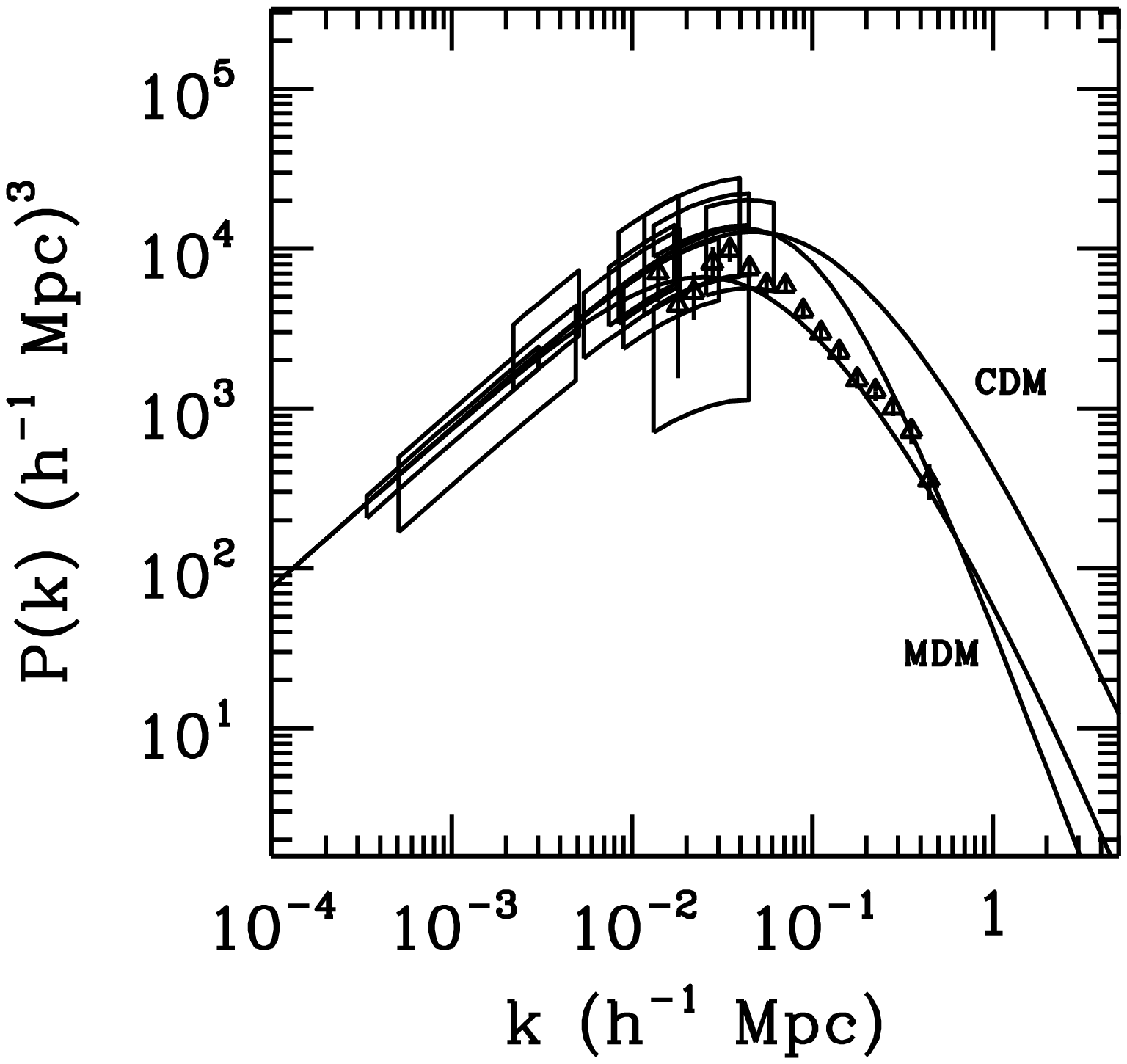}}
\caption{
{Comparison of the cold dark matter perturbation spectrum
with CBR anisotropy measurements (boxes) and the distribution
of galaxies today (triangles).  Wavenumber $k$ is
related to length scale, $k=2\pi /\lambda$; error flags
are not shown for the galaxy distribution.  The
curve labeled MDM is hot + cold dark matter (``5\,eV''
worth of neutrinos); the other two curves are cold dark
matter models with Hubble constants of $50\kms\Mpc$ (labeled
CDM) and $35\kms\Mpc$.  (Figure courtesy of M.~White.)}
  }
\label{fig:2}
\end{figure}

\section{INFLATIONARY THEORY}

\subsection{Generalities}

As successful as the big-bang cosmology it suffers from a dilemma
involving initial data.  Extrapolating back, one finds that
the Universe apparently began from a very special state:
A slightly inhomogeneous and very flat Robertson-Walker spacetime.
Collins and Hawking showed that the
set of initial data that evolve to a spacetime that is
as smooth and flat as ours is today of measure zero \cite{collins}.
(In the context of simple grand unified theories, the hot big bang
suffers from another serious problem:  the extreme overproduction of superheavy
magnetic monopoles; in fact, it was an attempt to solve the
monopole problem which led Guth to inflation.)

The cosmological appeal of inflation is its ability to lessen the dependence
of the present state of the Universe upon the initial state.  Two
elements are essential to doing this:  (1) accelerated (``superluminal'')
expansion and the concomitant tremendous growth of the
scale factor; and (2) massive entropy production \cite{htw}.
Together, these two features allow a small, smooth subhorizon-sized
patch of the early Universe to grow to a large enough size and contain enough
heat (entropy in excess of $10^{88}$) to easily encompass our
present Hubble volume.  Provided that
the region was originally small compared to the curvature
radius of the Universe it would appear flat then and today (just
as any small portion of the surface of a sphere appears flat).

While there is presently no standard model of inflation---just as
there is no standard model for physics at these energies
(typically $10^{15}\GeV$ or so)---viable models have much in
common.  They are based upon well posed, albeit highly speculative,
microphysics involving the classical evolution of a scalar field.
The superluminal expansion is driven by the potential
energy (``vacuum energy'') that arises when the scalar field
is displaced from its potential-energy minimum, which results in
nearly exponential expansion.  Provided
the potential is flat, during the time it takes for the field to roll
to the minimum of its potential the Universe undergoes many e-foldings
of expansion (more than around 60 or so are required to realize
the beneficial features of inflation).
As the scalar field nears the minimum,  the vacuum energy has been
converted to coherent oscillations of the scalar field, which
correspond to nonrelativistic scalar-field particles.  The eventual
decay of these particles into lighter particles and their thermalization
results in the ``reheating'' of the Universe and accounts for all the
heat in the Universe today (the entropy production event).

Superluminal expansion and the tremendous growth of the scale
factor (by a factor greater than that since the end of inflation)
allow quantum fluctuations on very small scales ($\la 10^{-23}\rcm$)
to be stretched to astrophysical scales ($\ga 10^{25}\rcm$).
Quantum fluctuations in the scalar field responsible for inflation
ultimately lead to an almost scale-invariant spectrum
of density perturbations \cite{scalar}, and quantum fluctuations in
the metric itself lead to an almost scale-invariant spectrum of gravity-waves
\cite{tensor}.   Scale invariance for density perturbations means
scale-independent fluctuations in the gravitational potential
(equivalently, density perturbations of different wavelength
cross the horizon with the same amplitude);
scale invariance for gravity waves means that
gravity waves of all wavelengths cross the horizon with the same amplitude.
Because of subsequent evolution, neither the scalar nor the
tensor perturbations are scale invariant today.

\subsection{Metaphysical implications}

Inflation alleviates the ``specialness'' problem greatly, but does
not eliminate all dependence upon the initial state \cite{nohair}.
All open FRW models will inflate and become flat; however,
many closed FRW models will recollapse before they can inflate.
If one imagines the most general initial spacetime as being comprised
of negatively and positively curved FRW (or Bianchi) models that are
stitched together, the failure of the positively
curved regions to inflate is of little consequence:  because
of exponential expansion during inflation the negatively curved
regions will occupy most of the space today.
Nor does inflation solve the smoothness problem forever;
it just postpones the problem into the exponentially distant future:
We will be able to see outside our smooth inflationary
patch and $\Omega$ will
start to deviate significantly from unity at a time $t\sim t_0
\exp [3(N-N_{\rm min}]$, where $N$ is the actual number of e-foldings
of inflation and $N_{\rm min}\sim 60$ is the minimum required to
solve the horizon/flatness problems.

Linde has emphasized that inflation has changed our view
of the Universe in a very fundamental way \cite{eternal}.
While cosmologists have long used
the Copernician principle to argue that the Universe must be smooth
because of the smoothness of our Hubble volume, in the post-inflation
view, our Hubble volume is smooth because it is a small
part of a region that underwent inflation.  On
the largest scales the structure of the Universe is likely to be
very rich:  Different regions may have undergone different amounts of
inflation, may have different laws of physics because they
evolved into different vacuum states (of equivalent energy), and
may even have different numbers of spatial dimensions.  Since it is
likely that most of the volume of the Universe is still undergoing
inflation and that inflationary patches are being constantly produced
(eternal inflation), the age of the Universe is
a meaningless concept and our expansion age merely measures the
time back to the end of our inflationary event!

\subsection{Specifics}

In Guth's seminal paper \cite{guth} he introduced the idea of inflation,
sung its praises, and showed that the model that he based the idea
upon did not work!  Thanks to very important contributions by Linde
\cite{linde} and Albrecht and Steinhardt \cite{as} that was quickly
remedied, and today there are many viable models of inflation.
That of course is both good news and bad news; it means that there
is no standard model of inflation.  Again, the absence of a standard
model of inflation should be viewed in the light of our general
ignorance about fundamental physics at these energies.

Many different approaches have taken in constructing particle-physics
models for inflation.  Some have focussed on very simple scalar potentials,
e.g., $V(\phi ) = \lambda \phi^4$ or $=m^2\phi^2/2$, without regard
to connecting the model to any underlying theory \cite{chaotic,pjsmst}.
Others have proposed more complicated models that attempt to make
contact with speculations about physics at very high energies,
e.g., grand unification \cite{pi}, supersymmetry \cite{florida,olive,lbl},
preonic physics \cite{pati}, or supergravity \cite{liddle}.
Several authors have attempted
to link inflation with superstring theory \cite{banks} or ``generic
predictions'' of superstring theory such as pseudo-Nambu-Goldstone
boson fields \cite{unnatural}.  While the scale of the vacuum energy
that drives inflation is typically of order $(10^{15}\GeV)^4$, a
model of inflation at the electroweak scale, vacuum energy $\approx(1\TeV )^4$,
has been proposed \cite{knox}.  There are also models in which there are
multiple epochs of inflation \cite{multiple}.

In all of the models above gravity is described by
general relativity.  A qualitatively different approach is to
consider inflation in the context of alternative theories of
gravity.  (After all, inflation probably involves physics at
energy scales not too different from the Planck scale and the
effective theory of gravity at these energies could well be
very different from general relativity; in fact, there are some
indications from superstring theory that gravity in these
circumstances might be described by a Brans-Dicke like theory.)
Perhaps the most successful of these models is first-order inflation
\cite{lapjs,kolbreview}.  First-order inflation returns
to Guth's original idea of a strongly first-order
phase transition; in the context of general relativity Guth's model
failed because the phase transition, if inflationary, never completed.
In theories where the effective strength of gravity evolves, like
Brans-Dicke theory, the weakening of gravity during inflation
allows the transition to complete.  In other models based upon
nonstandard gravitation theory, the scalar field responsible for
inflation is itself related to the size of additional spatial dimensions,
and inflation then also explains why our three spatial dimensions are
so big, while the other spatial dimensions are so small.

All models of inflation have one feature in common:  the scalar
field responsible for inflation has a very flat potential-energy
curve and is very weakly coupled.  This typically leads to a very
small dimensionless number, usually a dimensionless coupling
of the order of $10^{-14}$.
Such a small number, like other small numbers in physics (e.g.,
the ratio of the weak to Planck scales $\approx 10^{-17}$ or
the ratio of the mass of the electron to the $W/Z$ boson masses $\approx
10^{-5}$), runs counter to one's belief that a truly fundamental
theory should have no tiny parameters, and cries out for an
explanation.  At the very least, this small number must be stabilized against
quantum corrections---which it is in all of the previously
mentioned models.\footnote{It is sometimes stated that inflation
is unnatural because of the small coupling of the scalar field
responsible for inflation; while the small coupling certainly begs
explanation, these inflationary models are not unnatural in
the rigorous technical sense as the small number is stable
against quantum fluctuations.}  In some models, the small number in the
inflationary potential is related to other small numbers in
particle physics:  for example, the ratio of the electron mass
to the weak scale or the ratio of the unification scale to
the Planck scale.  Explaining the origin of
the small number that seems to be associated with
inflation is both a challenge and an opportunity.

Because of the growing base of observations that bear on inflation,
another approach to model building is emerging:  the use of observations
to constrain the underlying inflationary potential---and hence the
title of this paper.  In the next Section I focus on just this.
Before I do, I want to emphasize that while there are many
varieties of inflation, there are robust predictions
which are crucial to sharply testing inflation.

\subsection{Three robust predictions}

Inflation makes three robust\footnote{Because theorists
are so clever, it is not possible nor prudent to use the word
immutable.  Models that violate any or all of these ``robust
predications'' can and have been constructed.} predictions:

\begin{enumerate}

\item {\bf Flat universe.}  Because solving the ``horizon''
problem (large-scale smoothness in spite of small particle
horizons at early times) and solving the ``flatness'' problem
(maintaining $\Omega$ very close to unity until the present epoch)
are linked geometrically \cite{eu,htw}, this is the most robust
prediction of inflation.  Said another way, it is the prediction
that most inflationists would be least willing to give up.
(Even so, models of inflation have been
constructed where the amount of inflation is tuned just to give
$\Omega_0$ less than one today \cite{pu}.)  Through the Friedmann
equation for the scale factor, flat implies that the total
energy density (matter, radiation, vacuum energy, ...) is
equal to the critical density.

\item {\bf Nearly scale-invariant spectrum of gaussian density perturbations.}
Essentially all inflation models predict a nearly scale-invariant
spectrum of gaussian density perturbations.  Described in terms
of a power spectrum, $P(k) \equiv \langle |\delta_k|^2 \rangle
= Ak^n$, where $\delta_k$ is the Fourier transform of the primeval
density perturbations, and the spectral index $n\approx 1$ is equal to
unity in the scale-invariant limit.  The overall amplitude $A$
is model dependent.  Density perturbations give rise to CBR anisotropy
as well as seeding structure formation.
Requiring that the density perturbations are consistent
with the observed level of anisotropy of the CBR (and large enough
to produce the observed structure formation) is the most severe
constraint on inflationary models and leads to the small
dimensionless number that all inflationary models have.

\item {\bf Nearly scale-invariant spectrum of gravitational waves.}
These gravitational waves have wavelengths from ${\cal O}(1\km )$
to the size of the present Hubble radius and beyond.  Described in
terms of a power spectrum for the dimensionless gravity-wave amplitude
at early times, $P_T(k) \equiv \langle |h_k|^2 \rangle = A_Tk^{n_T-3}$, where
the spectral index $n_T \approx 0$ in the scale-invariant limit.
Once again, the overall amplitude $A_T$ is model dependent (varying
as the value of the inflationary vacuum energy).  Unlike density
perturbations, which are required to initiate structure formation,
there is no cosmological lower bound to the amplitude of
the gravity-wave perturbations.  Tensor perturbations
also give rise to CBR anisotropy; requiring that they do not lead to
excessive anisotropy implies that the energy density that drove
inflation must be less than about $(10^{16}\GeV )^4$.  This
indicates that if inflation took place, it did so at an energy well
below the Planck scale.\footnote{To be more precise, the part of inflation
that led to perturbations on scales within the present horizon involved
subPlanckian energy densities.  In
some models of inflation, the earliest stages, which do not influence
scales that we are privy to, involve energies as large as the Planck scale.}

\end{enumerate}

There are other interesting consequences of inflation that
are less generic.  For example, in models of first-order inflation,
in which reheating occurs through the nucleation and collision of
vacuum bubbles, there is an additional, larger amplitude, but
narrow-band, spectrum of gravitational waves ($\Omega_{\rm GW}h^2
\sim 10^{-6}$) \cite{vacuumpop}.  In other models large-scale primeval magnetic
fields of interesting size are seeded during inflation \cite{bfield}.

\section{STRUCTURE FORMATION:  A WINDOW TO THE EARLY UNIVERSE}

The key to testing inflation is to focus on its robust predictions
and their implications.  Earlier I discussed the prediction of
a flat Universe and its bold implication that most of the matter in Universe
exists in the form of particle dark matter.  Much effort is being
directed at determining the mean density of the Universe and
detecting particle dark matter.

The scale-invariant scalar metric perturbations
lead to CBR anisotropy on angular scales from less
than $1^\circ$ to $90^\circ$
and seed the formation of structure in the Universe.
Together with the nucleosynthesis determination of $\Omega_B$ and
the inflationary prediction of a flat Universe, scale-invariant
density perturbations lead to a very specific scenario for structure
formation; it is known
as cold dark matter because the bulk of the particle dark
matter is comprised of slowly moving particles (e.g., axions or
neutralinos) \cite{cdmrev}.\footnote{The simpler possibility,
that the particle dark matter exists in the form of $30\eV$ or so
neutrinos which is known as hot dark matter, was falsified almost
a decade ago.  Because neutrinos move rapidly, they can diffuse
from high density to low density regions damping perturbations on
small scales.  In hot dark matter large, supercluster-size objects
must form before galaxies, and thus hot dark matter cannot account
for the abundance of galaxies, damped Lyman-$\alpha$ clouds, etc.
that is observed at high redshift.}   A large and rapidly growing number
of observations are being brought to bear in the testing of
cold dark matter, making it the centerpiece of efforts to test inflation.

Finally, there are the scale-invariant tensor perturbations.  They
lead to CBR anisotropy on angular scales from a few degrees to
$90^\circ$ and a spectrum of gravitational waves.  The CBR anisotropy
arising from the tensor perturbations can in principle be separated
from that arising from scalar perturbations.  However, because the sky is
finite, sampling variance sets a fundamental limit:  the tensor
contribution to CBR anisotropy can only be separated from that of
the scalar if $T/S$ is greater than about $0.14$ \cite{knoxmst}
($T$ is the contribution of tensor perturbations to the variance
of the CBR quadrupole and $S$ is the same for scalar perturbations).  It is
also possible that the stochastic background of gravitational
waves itself can be directly detected, though it appears that
the LIGO facilities being built will lack the sensitivity and
even space-based interferometery (e.g., LISA) is not a sure bet \cite{tlw}.

Before going on to discuss how cold dark matter models are testing
inflation I want to emphasize the importance
of the tensor perturbations.   The attractiveness of a flat Universe
with scale-invariant density perturbations was appreciated long before
inflation.  Verifying these two predictions of inflation, while
important, will not provide a ``smoking gun.''  The tensor perturbations
are a unique feature of inflation.  Further, they are crucial
to obtaining information about the scalar potential responsible for inflation.

\subsection{Vanilla Cold Dark Matter: almost, but not quite?}

The simplest version of cold dark matter, vanilla cold dark matter
if you will, is characterized by:  (1) $\Omega_B \sim 0.5$
and $\Omega_{\rm CDM} \sim 0.95$; (2) Hubble constant of $50\kms\Mpc^{-1}$;
(3) Precisely scale-invariant density perturbations ($n=1$); and
(4) No contribution of tensor perturbations to CBR anisotropy.
In cold dark matter models structure forms hierarchically, with small objects
forming first and merging to form larger objects.  Galaxies form
at redshifts of order a few, and rarer objects like QSOs form from higher
than average density peaks earlier.
In general, cold dark matter predicts a Universe that is still
evolving at recent epochs.  $N$-body simulations
are crucial to bridging the gap between theory and observation,
and several groups have carried out large numerical studies of
vanilla cold dark matter \cite{nbody}.

There are a diversity of observations that test cold dark matter; they
include CBR anisotropy and spectral distortions, redshift surveys, pairwise
velocities of galaxies, peculiar velocities,
redshift space distortions, x-ray background,
QSO absorption line systems, cluster studies of all kinds,
studies of evolution (clusters, galaxies, and so on), measurements
of the Hubble constant, and on and on.  I will focus on how these
measurements probe the power spectrum of density perturbations,
emphasizing the role of
CBR-anisotropy measurements and redshift surveys.

Density perturbations on a (comoving) length scale $\lambda$
give rise to CBR anisotropy on an angular scale $\theta \sim
\lambda/H_0^{-1} \sim 1^\circ (\lambda /100h^{-1}\Mpc )$.\footnote{For
reference, perturbations on a length scale of about $1\Mpc$ give
rise to galaxies, on about $10\Mpc$ to clusters, on about $30\Mpc$
to large voids, and on about $100\Mpc$ to the great walls.}
CBR anisotropy has now been detected by more than ten experiments
on angular scales from about $0.5^\circ$ to $90^\circ$, thereby
probing length scales from $30h^{-1}\Mpc$ to $10^4h^{-1}\Mpc$.
The very accurate measurements made by the COBE DMR can be used
to normalize the cold dark matter spectrum (the normalization
scale corresponds to about $20^\circ$).  When this is done, the
other ten or so measurements are in agreement with the predictions
of cold dark matter (see Fig.~1).

The COBE-normalized cold dark matter spectrum can be extrapolated
to the much smaller scales probed by redshift surveys, from about
$1h^{-1}\Mpc$ to $100h^{-1}\Mpc$.  When this is done, there is general
agreement.  However, on closer inspection the COBE-normalized spectrum
seems to predict excess power on these scales (about a factor
of four in the power spectrum; see Fig.~2).  This conclusion
is supported by other observations, e.g., the abundance of rich
clusters and the pairwise velocities of galaxies.  It suggests
that cold dark matter has much of the truth, but perhaps not
all of it \cite{cdmproblems}, and has led to the suggestion that
something needs to be added to the simplest cold dark matter theory.

There is another important challenge facing cold dark matter.
X-ray observations of rich clusters are able to determine the ratio
of hot gas (baryons) to total cluster mass (baryons + CDM) (by a wide
margin, most of the baryons ``seen'' in clusters are in the hot gas).
To be sure there are assumptions and uncertainties; the data at the moment
indicate that this ratio is $0.04h^{-3/2}-0.1h^{-3/2}$ \cite{clusters}.
If clusters provide a fair sample of the universal mix of matter,
then this ratio should equal $\Omega_B/(\Omega_B + \Omega_{\rm CDM})
\simeq (0.009-0.022)h^{-2}/ (\Omega_B + \Omega_{\rm CDM})$.  Since
clusters are large objects they should provide a pretty fair sample.
Taking the numbers at face value, cold dark matter is consistent with the
cluster gas fraction provided either:  $\Omega_B + \Omega_{\rm CDM}
= 1$ and $h\sim 0.3$ or $\Omega_B + \Omega_{\rm CDM} \sim 0.3$ and
$h\sim 0.7$.  The cluster baryon problem has yet to be settled,
and is clearly an important test of cold dark matter.
                                                       
Finally, before going on to discuss the variants of cold dark matter now under
consideration, let me add a note of caution.
The comparison of predictions for structure formation
with present-day observations of the
distribution of galaxies is fraught with difficulties.
Theory most accurately predicts ``where the mass is''
(in a statistical sense) and the observations determine where the light is.
Redshift surveys probe present-day inhomogeneity on scales
from around one $\Mpc$ to a few hundred $\Mpc$, scales where
the Universe is nonlinear ($\delta n_{\rm GAL}/n_{\rm GAL}
\ga 1$ on scales $\la 8h^{-1}\Mpc$) and where astrophysical
processes undoubtedly play an important role
(e.g., star formation determines where and when
``mass lights up,'' the explosive release of energy in supernovae
can move matter around and influence subsequent star formation,
and so on).  The distance to a galaxy is
determined through Hubble's law ($d = H_0^{-1} z$) by measuring a redshift;
peculiar velocities induced by the lumpy distribution
of matter are significant and prevent a direct determination
of the actual distance.  There are the intrinsic limitations
of the surveys themselves:  they are flux not volume limited (brighter
objects are seen to greater distances and vice versa) and relatively
small (e.g., the CfA slices of the Universe survey contains only
about $10^4$ galaxies and extends to a redshift of about $z\sim 0.03$).
Last but not least are the numerical
simulations which link theory and observation;
they are limited in dynamical range (about a factor
of 100 in length scale) and in microphysics (in the largest simulations
only gravity, and in others only a gross approximation to the effects of
hydrodynamics/thermodynamics).  Perhaps it would be prudent to withhold
judgment on vanilla cold dark matter for the moment and resist
the urge to modify it---but that wouldn't be as much fun!

\subsection{The many flavors of cold dark matter}

The spectrum of density perturbations today depends
not only upon the primeval spectrum (and the normalization
on large scales provided by COBE), but also upon the energy content
of the Universe.  While the fluctuations in the gravitational potential
were initially (approximately) scale invariant,
the Universe evolved from an early radiation-dominated phase
to a matter-dominated phase which imposes a characteristic scale
on the spectrum of density perturbations seen today;
that scale is determined by the energy
content of the Universe, $k_{\rm EQ}\sim 10^{-1}h\Mpc^{-1}
\,(\Omega_{\rm matter}h/\sqrt{g_*})$
($g_*$ counts the relativistic degrees of freedom,
$\Omega_{\rm matter} = \Omega_B +\Omega_{\rm CDM}$).
In addition, if some of the nonbaryonic dark
matter is neutrinos, they reduce power on small
scales somewhat through freestreaming (see Fig.~2).
With this in mind, let me discuss
the variants of cold dark matter that have been proposed to
improve its agreement with observations.

\begin{enumerate}

\item {\bf Low Hubble Constant + cold dark matter (LHC CDM) \cite{lhc}.}
Remarkably, simply lowering the Hubble constant to around
$30\kms\Mpc^{-1}$ solves all the problems
of cold dark matter.  Recall, the critical density $\rho_{\rm crit}
\propto H_0^2$; lowering $H_0$ lowers the matter density and
has precisely the desired effect.  It has two other added benefits:
the expansion age of the Universe is comfortably
consistent with the ages of the
oldest stars and the baryon fraction is raised to a value that is
consistent with that measured in x-ray clusters.  Needless to say, such a
small value for the Hubble constant flies in the face of current
observations \cite{hubble}; further, it illustrates that the problems of
cold dark matter get even worse for the larger values of $H_0$
that are favored by recent observations.

\item {\bf Hot + cold dark matter ($\nu$CDM) \cite{chdm}.}
Adding a small amount of
hot dark matter can suppress density perturbations on small scales;
adding too much leads back to the longstanding
problems of hot dark matter.  Retaining enough power
on very small scales to produce damped Lyman-$\alpha$ systems at
high redshift limits $\Omega_\nu$ to less than about
20\%, corresponding to about ``$5\eV$ worth
of neutrinos'' (i.e., one species of mass $5\eV$, or two species
of mass $2.5\eV$, and so on).  This admixture of hot dark matter
rejuvenates cold dark matter provided the Hubble constant is not
too large, $H_0\la 55 \kms\Mpc^{-1}$; in fact, a Hubble constant
of closer to $45\kms\Mpc^{-1}$ is preferred.

\item {\bf Cosmological constant + cold dark matter ($\Lambda$CDM) \cite{lcdm}.}
(A cosmological constant corresponds to a uniform energy density,
or vacuum energy.)  Shifting 50\% to 80\% of the critical density
to a cosmological constant lowers the matter density
and has the same beneficial effect as a low Hubble constant.
In fact, a Hubble constant as large as $80\kms\Mpc^{-1}$
can be accommodated.  In addition,
the cosmological constant allows the age problem to solved
even if the Hubble constant is large, addresses the fact
that few measurements of the mean mass density give a value as large as
the critical density (most measurements of the mass density
are insensitive to a uniform component), and allows the
baryon fraction of matter to be larger, which
alleviates the cluster baryon problem.   Not everything is rosy;
cosmologists have invoked a cosmological constant twice before to solve
their problems (Einstein to obtain a static universe and
Bondi, Gold, and Hoyle to solve the earlier age crisis when
$H_0$ was thought to be $250\kms\Mpc^{-1}$).  Further, particle
physicists can still not explain why the energy of the
vacuum is not at least 50 (if not 120) orders of magnitude larger than
the present critical density, and expect that when the problem is
solved the answer will be zero.

\item {\bf Extra relativistic particles + cold
dark matter ($\tau$CDM) \cite{taucdm}.}
Raising the level of radiation has the same beneficial effect
as lowering the matter density.  In the standard cosmology the
radiation content consists of photons + three (undetected)
cosmic seas of neutrinos (corresponding to $g_* \simeq
3.36$).   While we have no direct determination
of the radiation beyond that in the CBR, there are
at least two problems:  What are the additional relativistic
particles? and Can additional radiation be added without
upsetting the successful predictions
of primordial nucleosynthesis which depend critically upon the
energy density of relativistic particles?  The simplest way around
these problems is an unstable tau neutrino (mass anywhere
between a few keV and a few MeV) whose decays produce
the radiation.   This fix can tolerate a larger Hubble constant,
though at the expense of more radiation.

\item {\bf Tilted cold dark matter (TCDM) \cite{tcdm}.}
While the spectrum of density
perturbations in most models of inflation is very nearly scale invariant,
there are models where the deviations are significant ($n\approx 0.8$)
which leads to smaller fluctuations on small scales.  Further,
if gravity waves account for a significant
part of the CBR anisotropy, the level of density perturbations can be
lowered even more.  A combination of tilt and gravity waves can solve the
problem of too much power on small scales, but seems to lead to
too little power on intermediate and very small scales.

\end{enumerate}

In evaluating these better fit models, one should keep the words
of Francis Crick in mind (loosely paraphrased):  A model that fits
all the data at a given time is necessarily wrong, because at any given
time not all the data are correct(!).  $\Lambda$CDM provides an
interesting/confusing example.  When I discussed it in 1990,
I called it the best-fit Universe, and quoting Crick, I said that
$\Lambda$CDM was certain to fall by the wayside \cite{lcdm0}.
In 1995, it is still the best-fit model \cite{lcdm1}.

Let me end by defending the other point of view, namely,
that to add something to cold dark matter is not unreasonable,
or even as some have said, a last gasp effort to saving a dying theory.
Standard cold dark matter was a starting point, similar to
early calculations of big-bang nucleosynthesis.  It was always
appreciated that the inflationary spectrum of density perturbations
was not exactly scale invariant \cite{pjsmst} and that the Hubble constant
was unlikely to be exactly $50\kms \Mpc$.  As the quality and quantity
of data improve, it is only sensible to refine the model, just as has been
done with big-bang nucleosynthesis.  Cold dark matter seems to
embody much of the ``truth.''   The modifications suggested
all seem quite reasonable (as opposed
to contrived).  Neutrinos exist; they are expected to have mass;
there is even some experimental data that indicates they do have
mass.  It is still within the realm of possibility that the Hubble
constant is less than $50\kms\Mpc^{-1}$, and if it is as large
as $70\kms\Mpc^{-1}$ to $80\kms\Mpc^{-1}$ a cosmological constant
seems inescapable based upon the age problem alone.  There is no data
that can preclude more radiation than in the standard cosmology
and deviations from scale invariance were always expected.

\subsection{Reconstruction}

If inflation and the cold dark matter theory is shown to be correct,
then a window to the very early Universe ($t\sim 10^{-34}\sec$) will
have been opened.  While it is certainly premature to jump to this
conclusion, I would like to illustrate one example of what one could
hope to learn.  As mentioned earlier, the spectra and amplitudes
of the the tensor and scalar metric perturbations predicted by
inflation depend upon the underlying model, to be specific, the
shape of the inflationary scalar-field potential.
If one can measure the power-law
index of the scalar spectrum and the amplitudes of the scalar
and tensor spectra, one can recover the value of the potential
and its first two derivatives around the point on the potential
where inflation took place \cite{reconstruct}:
\begin{eqnarray}
V & = & 1.65 T\, {m_{\rm Pl}}^4  , \\
V^\prime & = & \pm \sqrt{8\pi r \over 7}\, V/{m_{\rm Pl}} , \\
V^{\prime\prime} & = & 4\pi \left[ (n-1) + {3\over 7} r \right]\,
V /{m_{\rm Pl}}^2 ,
\end{eqnarray}
where $r\equiv T/S$, a prime indicates derivative
with respect to $\phi$, $\mpl = 1.22\times 10^{19}\GeV$ is
the Planck energy, and the sign of $V^\prime$ is
indeterminate.  In addition, if the tensor spectral index
can be measured a consistency relation, $n_T = -r/7$,
can be used to further test inflation.
Reconstruction of the inflationary scalar potential would
shed light both on inflation as well as physics at energies of the
order of $10^{15}\GeV$.

\section{The Future}

%pbi and texture

The stakes for cosmology are high:  if correct, inflation/cold dark matter
represents a major extension of the big bang and our understanding
of the Universe, which can't help but shed
light on the fundamental physics at energies of order $10^{15}\GeV$.

What are the crucial tests and when will they be carried out?
Because of the many measurements/observations
that can have significant impact, I believe the answer to when
is sooner rather than later.  The list of pivotal observations is long:
CBR anisotropy, large redshift surveys (e.g., the Sloan Digital
Sky Survey will have $10^6$ redshifts), direct searches
for nonbaryonic in our neighborhood (both for axions and neutralinos)
and baryonic dark matter (microlensing), x-ray studies of galaxy
clusters, the use of back-lit gas clouds (quasar absorption line systems)
to study the Universe at high redshift, evolution (as
revealed by deep images of the sky taken by the Hubble Space Telescope and
the Keck 10 meter telescope), measurements of both $H_0$ and $q_0$,
mapping of the peculiar velocity field at large redshifts
through the Sunyaev-Zel'dovich effect, dynamical estimates of
the mass density (using weak gravitational lensing, large-scale velocity
fields, and so on), age determinations,
gravitational lensing, searches for supersymmetric
particles (at accelerators) and neutrino oscillations (at accelerators,
solar-neutrino detectors, and other large underground detectors),
searches for high-energy neutrinos from
neutralino annihilations in the sun using large underground detectors,
and on and on.  Let me end by illustrating the interesting
consequences of several possible measurements.

A definitive determination that $H_0$ is greater than $55\kms\Mpc^{-1}$
would falsify LHC CDM and $\nu$CDM.  A definitive determination
that $H_0$ is $75\kms \Mpc^{-1}$ or larger would necessitate a
cosmological constant.  A flat Universe
with a cosmological constant has a very different deceleration
parameter than one dominated by matter, $q_0 =-1.5\Omega_\Lambda + 0.5
\sim -(0.4 - 0.7)$ compared to $q_0 = 0.5$, and this could be settled
by galaxy number counts or numbers of lensed quasars.
The level of CBR anisotropy in $\tau$CDM and LHC CDM on the $0.5^\circ$ scale
is about 50\% larger than the other models, which should be
easily discernible.   If neutrino-oscillation
experiments were to provide evidence for a neutrino of mass $5\eV$ (or two of
mass $2.5\eV$) $\nu$CDM would seem almost inescapable.

Many more CBR measurements are in progress and there should
many interesting results in the next few years.  In the wake
of the success of COBE there are proposals, both in the US and
Europe, for a satellite-borne instrument to map the CBR sky
with a factor of ten better resolution.
A map of the CBR with $0.5^\circ -1^\circ$ resolution could separate
the gravity-wave contribution to
CBR anisotropy and provide evidence for the third robust
prediction of inflation, as well as determining other important
parameters \cite{knoxphd}, e.g., the scalar and tensor indices,
$\Omega_\Lambda$, and even $\Omega_0$ (the position of
the ``Doppler'' peak scales as $0.5^\circ /\sqrt{\Omega_0}$)
\cite{omega}).

\section*{Acknowledgments}
This work was supported in part by the DOE (at Chicago and
Fermilab) and the NASA (at Fermilab through grant NAG 5-2788).

\end{document}